\documentclass[a4paper,12pt,english]{iopart}
\usepackage[T1]{fontenc}
\usepackage[utf8]{inputenc}

\usepackage[usenames,dvipsnames]{xcolor}
\usepackage{graphicx}
\usepackage{iopams}
\usepackage{setstack}
\usepackage{multirow}
\usepackage{siunitx}

\usepackage{cite}
\usepackage{babel}

\usepackage{epstopdf}
\usepackage{subfigure}
\usepackage{capt-of}

\begin{document}

\title[PLE spectroscopy of SiV$^{-}$ and GeV$^{-}$ color center in diamond]{Photoluminescence excitation spectroscopy of SiV$^{-}$ and GeV$^{-}$ color center in diamond}

\author{
Stefan Häußler$^{1,2}$,
Gergő Thiering$^3$,
Andreas Dietrich$^1$,
Niklas Waasem$^4$,
Tokuyuki Teraji$^5$,
Junichi Isoya$^6$,
Takayuki Iwasaki$^7$,
Mutsuko Hatano$^7$,
Fedor Jelezko$^{1,2}$,
Adam Gali$^3$ and
Alexander Kubanek$^{1,2}$
}

\address{$^1$ Institute for Quantum Optics, Ulm University, Albert-Einstein-Allee 11, D-89081 Ulm, Germany}
\address{$^2$ Center for Integrated Quantum Science and Technology (IQst), Ulm University, Albert-Einstein-Allee 11, D-89081 Ulm, Germany}
\address{$^3$ Institute for Solid State Physics and Optics, Wigner Research Centre for Physics, Hungarian Academy of Sciences, PO Box 49, H-1525 Budapest, Hungary}
\address{$^4$ Hübner GmbH \& Co. KG, Heinrich-Hertz-Straße 2, D-34123 Kassel, Germany}
\address{$^5$ National Institute for Materials Science, 1-1 Namiki, Tsukuba, Ibaraki 305-0044, Japan}
\address{$^6$ Research Center for Knowledge Communities, University of Tsukuba, 1-2 Kasuga, Tsukuba, Ibaraki 305-8550, Japan}
\address{$^7$ Department of Electrical and Electronic Engineering, Tokyo Institute of Technology, 2-12-1 Ookayama, Meguro-ku, Tokyo 152-8552, Japan}

\ead{stefan.haeussler@uni-ulm.de, alexander.kubanek@uni-ulm.de}

\begin{abstract}
 
Color centers in diamond are important quantum emitters for a broad range of applications ranging from quantum sensing to quantum optics. Understanding the internal energy level structure is of fundamental importance for future applications. We experimentally investigate the level structure of an ensemble of few negatively charged silicon-vacancy (SiV$^{-}$) and germanium-vacancy (GeV$^{-}$) centers in bulk diamond at room temperature by photoluminescence (PL) and excitation (PLE) spectroscopy over a broad wavelength range from $460 \, \text{nm}$ to $650 \, \text{nm}$ and perform power-dependent saturation measurements. For SiV$^{-}$ our experimental results confirm the presence of a higher energy transition at $\sim \! 2.31 \, \text{eV}$. By comparison with detailed theoretical simulations of the imaginary dielectric function we interpret the transition as a dipole-allowed transition from $^2E_{g}$-state to $^2A_{2u}$-state where the corresponding $a_{2u}$-level lies deeply inside the diamond valence band. Therefore, the transition is broadened by the diamond band. At higher excitation power of $10 \, \text{mW}$ we indicate signs of  a parity-conserving transition at $\sim \! 2.03 \, \text{eV}$ supported by saturation measurements. For GeV$^{-}$ we demonstrate that the PLE spectrum is in good agreement with the mirror image of the PL spectrum of the zero-phonon line (ZPL). Experimentally we do not observe a higher lying energy level up to a transition wavelength of $460 \, \text{nm}$. The observed PL spectra are identical, independent of excitation wavelength, suggesting a rapid decay to $^2E_{u}$ excited state and followed by optical transition to $^2E_{g}$ ground state. Our investigations convey important insights for future quantum optics and quantum sensing experiments based on SiV$^{-}$-center and GeV$^{-}$-center in diamond. 

\end{abstract}


\maketitle

\section{Introduction}

Color centers in diamond are used in a large set of applications including luminescent markers \cite{aharonovich2011diamond}, magnetic field sensing with high spatial resolution \cite{balasubramanian2009ultralong} or as single photon emitter in quantum optics applications \cite{togan2010quantum,bernien2013heralded}. Diamond is of particular interest since it can accommodate a large number of different color centers giving access to a large spectral range and variety of different spin and optical properties. The negatively charged nitrogen-vacancy center (NV$^{-}$) is the most studied color center with a unique set of properties and applications \cite{wrachtrup2006single}. However, the search for new color centers is of great importance not only to extend the spectral range but also to find color centers with improved spin and, in particular, optical properties. Emission of single photons has been demonstrated with Cr-based color centers \cite{aharonovich2010chromium} and Ni-based centers (NE8) \cite{gaebel2004stable}, but engineering of them is difficult. \\
Recently the SiV$^{-}$ center has drawn great attention due to its unique set of optical properties \cite{muller2014optical} demonstrating a comparably large Debye-Waller (DW) factor of $\sim \! 0.7$, single photon emission with good polarization contrast \cite{rogers2014electronic}, a very small inhomogeneous line broadening of only a few transition linewidth and a large spectral stability enabling a spectral overlap of up to $91 \, \%$ and lifetime-limited linewidth at resonant excitation \cite{rogers2014multiple}. These developments enabled the demonstration of indistinguishable, single photon emission employing Hong-Ou-Mandel interference without the need of electric field tuning of the transition frequency \cite{sipahigil2014indistinguishable}. The $D_{3d}$ symmetry of the SiV$^{-}$ center plays a crucial role for the spectral stability, leading to a very small sensitivity to electric fields and, correspondingly, a small spectral diffusion and narrow inhomogeneous distribution of transition frequencies. \\
The GeV$^{-}$ center was proposed to offer a very similar structure to that of SiV$^{-}$ \cite{goss2005vacancy}. Very recently Ge-related color centers were reported in high-pressure-high-temperature (HPHT) synthesized diamond crystals \cite{ekimov2015germanium,palyanov2015germanium} experimentally demonstrating zero-phonon line (ZPL) emission at $602.5 \, \text{nm}$ with a full width at half maximum (FWHM) of $4.5 \, \text{nm}$ at room temperature. From DFT calculations the electron excitation energy for the GeV$^{-}$ center was calculated to be $2.01 \, \text{eV}$. Single photon emission at $602 \, \text{nm}$ was demonstrated for GeV$^{-}$ centers produced from ion implantation \cite{iwasaki2015germanium}. \\
A key requirement for future applications is the understanding of the internal energy level structure with its excitation and luminescence properties. While for GeV$^{-}$ the energy level structure, in particular, of higher-lying states has so far barely been studied, first measurements with SiV centers have been reported in 2000 \cite{iakoubovskii2000photochromism} recording photoluminescence excitation (PLE) spectra over a large excitation wavelength range from $490 \, \text{nm}$ to $690 \, \text{nm}$ employing a $250 \, \text{W}$ halogen lamp in combination with double monochromator therefore measuring with low spectral power density of the excitation light. \\
In this paper we investigate PLE spectroscopy for both, an ensemble of few SiV$^{-}$ centers and an ensemble of few GeV$^{-}$ centers in bulk diamond at room temperature. We measure PLE spectra over an excitation wavelength range from $460 \, \text{nm}$ to $650 \, \text{nm}$ and examine the power-dependent saturation for each measurement. We compare our experimental results with theoretical simulations of the imaginary dielectric function and develop a consistent physical picture. 

\cleardoublepage 
\newpage
\newpage

\section{Methods}

Our experiments are performed using a custom-built confocal microscope with a high NA air objective (NA = $0.9$). As a tunable laserlight source, a continuous-wave (cw) optical-parametric oscillator (OPO) with second-harmonic-generation unit (SHG) is employed (C-WAVE from Hübner). The system covers the wavelength ranges $900 - 1300 \, \text{nm}$ (OPO) and $450 - 650 \, \text{nm}$ (SHG) with output powers in the range of several hundred milliwatts. The linewidth of the cw output beam is below $1 \, \text{MHz}$ and the frequency can be swept mode-hop free over $25 \, \text{GHz}$. The excitation power in the sample plane reaches up to $20 \, \text{mW}$. The photoluminescence (PL) of the different color centers is cleaned up using several spectral filters in the detection path and measured either using a single-photon counting module (SPCM) or a spectrometer with a 150 groves/mm grating. The samples are mounted on a two axis piezo nanopositioning stage enabling mapping with scan ranges of up to $200 \, \mu\text{m}^{2}$. \\
We apply \emph{ab initio} theoretical modeling, in order to characterize the optical properties of the negatively charged SiV and GeV defects and compare them with the experimental observations. We employ spin-polarized density functional theory (DFT) plane-wave supercell calculations as implemented in the \textsc{vasp} code \cite{Kresse_1994, Kresse_1996}. We employ projector augmented-wave\cite{Blochl_1994} (PAW) method to treat the ions with $370 \, \text{eV}$ plane wave cutoff for the pseudo wavefunctions, and relax the geometries until the forces acting on the ions fall below $10^{-2} \, \text{eV/\AA}$. A small-core projector for the Ge ion is applied in the calculations. We use different techniques in order to calculate the optical spectrum of the defects. The excitation energies are calculated within the range separated hybrid density functional HSE06 \cite{Jochen_2003, Aliaksandr_2006} with applying the constraint DFT method \cite{Gali_2009, Deak_2010}. We employ simple cubic 512-atom diamond supercell with $\Gamma$-point sampling of the Brillouin zone in these calculations. The band to defect level optical transitions are estimated within DFT as more sophisticated methods such as Bethe-Salpeter equation are computationally intractable. The considered defects have quasi $D_{3d}$ symmetry with $S=1/2$ spin state with a quasi-degenerate half-occupied in-gap defect level in the spin minority channel. This is a complicated situation where the band to defect level optical transition cannot be readily calculated by using high number of k-points in the Brillouin zone, as the dispersion of the defect levels in the gap would result in false occupation of the in-gap states that leads to artificial optical transitions. Instead, we increase the size of the supercell with keeping the $\Gamma$-point sampling of the Brillouin-zone. We apply cubic diamond supercells up to 4096-atom in the calculation of absorption spectrum which is taken from the trace of the imaginary part of the dielectric function \cite{Gajdos_2006}. This type of calculations are extremely demanding even at simple Kohn-Sham DFT level. Thus, we apply the computationally less expensive PBE \cite{Perdew_1996} functional in these calculations. In our experience, the PBE and HSE06 Kohn-Sham wavefunctions are very similar for these defects, thus the transition dipole moments are well approximated by PBE calculations. We note that 4096-atom supercell is still not entirely convergent, nevertheless, it provides a semi-quantitative estimate.  

\cleardoublepage 
\newpage
\newpage

\section{Investigation of ensemble of few SiV$^{-}$ center}

\begin{figure}[htbp]
	\begin{minipage}{0.49\textwidth}
			\textbf{(a)} \\
			\includegraphics[width=\textwidth]{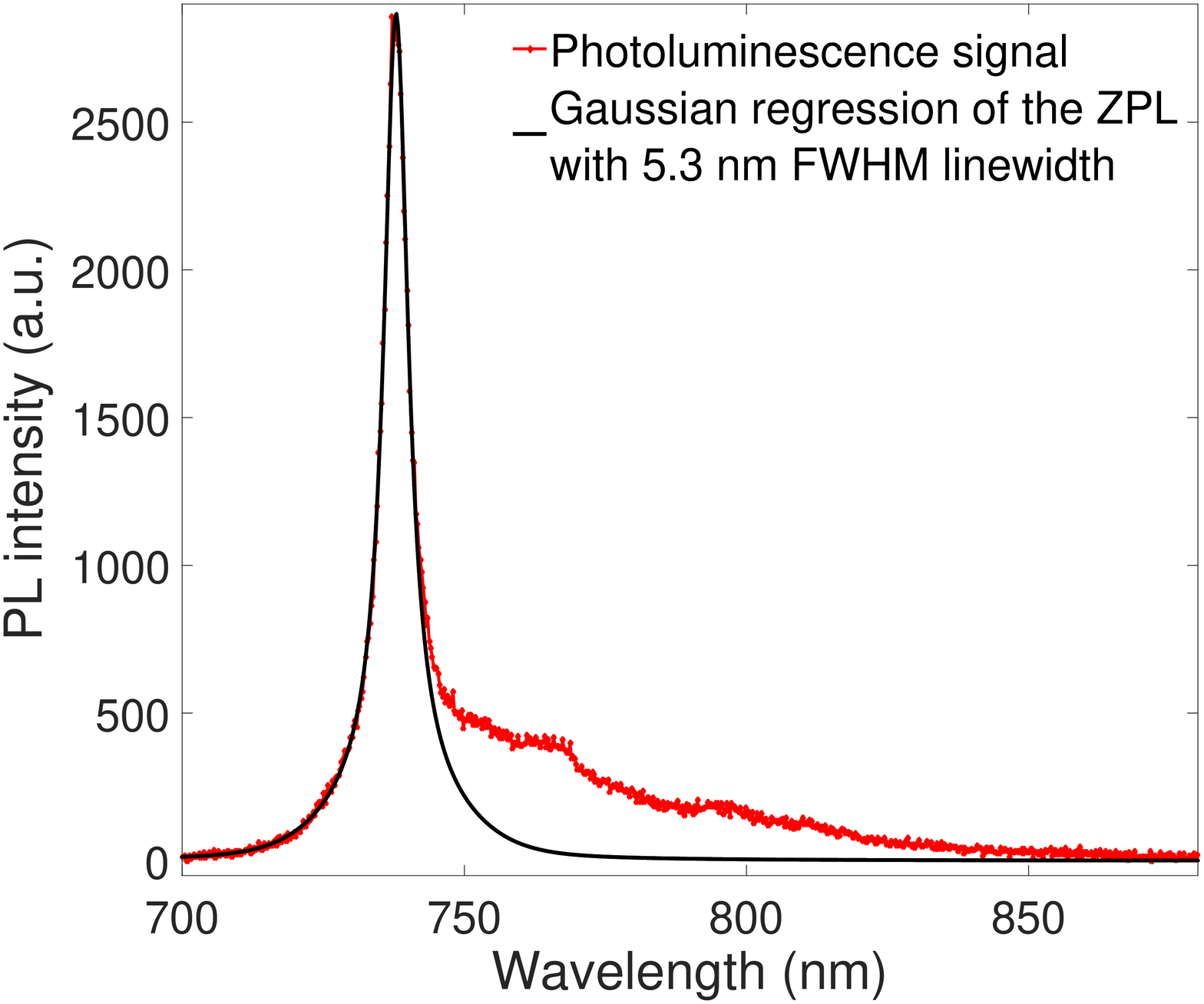}
	\end{minipage}
	\begin{minipage}{0.49\textwidth}
			\textbf{(c)} \\
			\includegraphics[width=\textwidth]{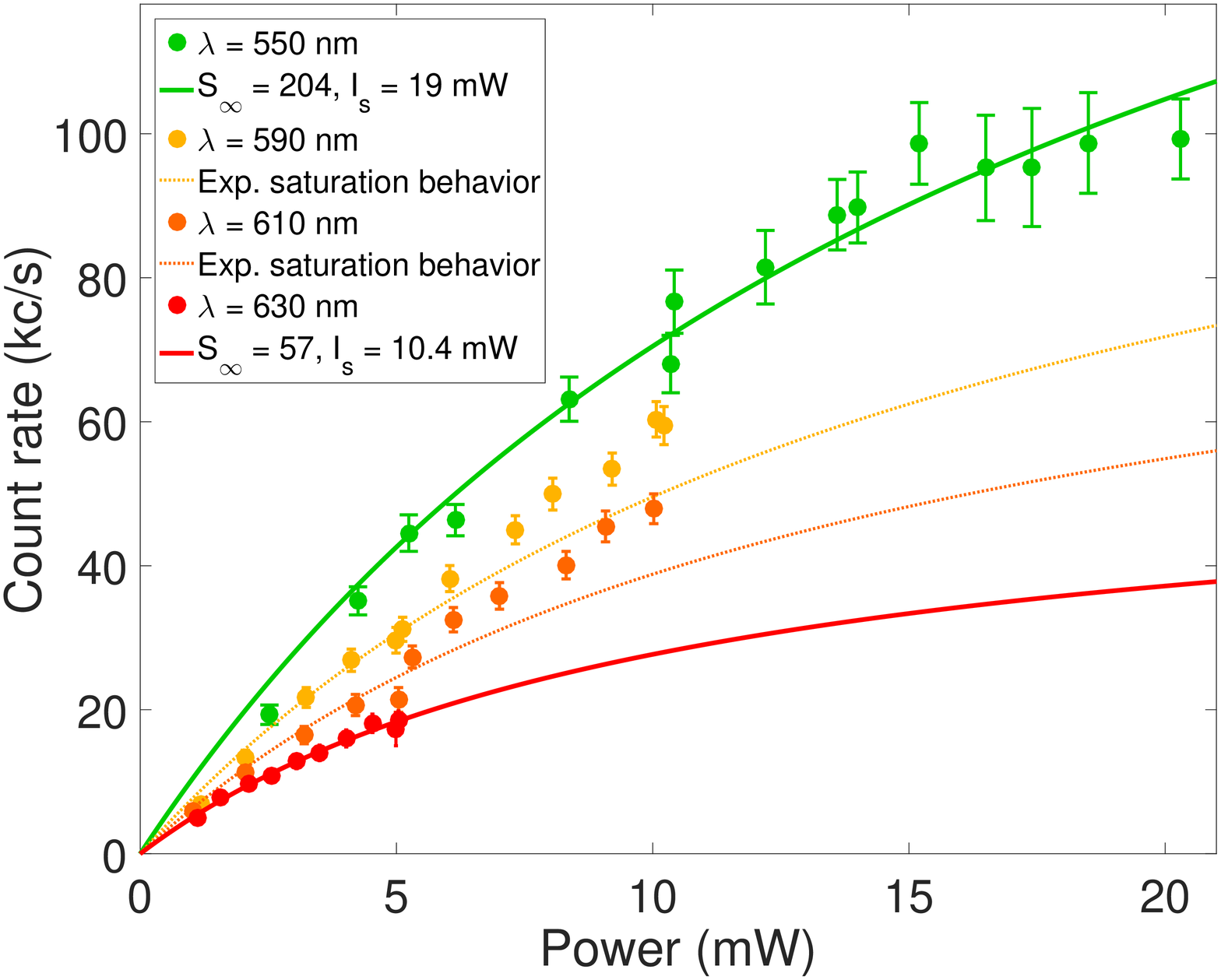}
	\end{minipage}
	\begin{minipage}{\textwidth}
			\textbf{(b)} \\
			\includegraphics[width=\textwidth]{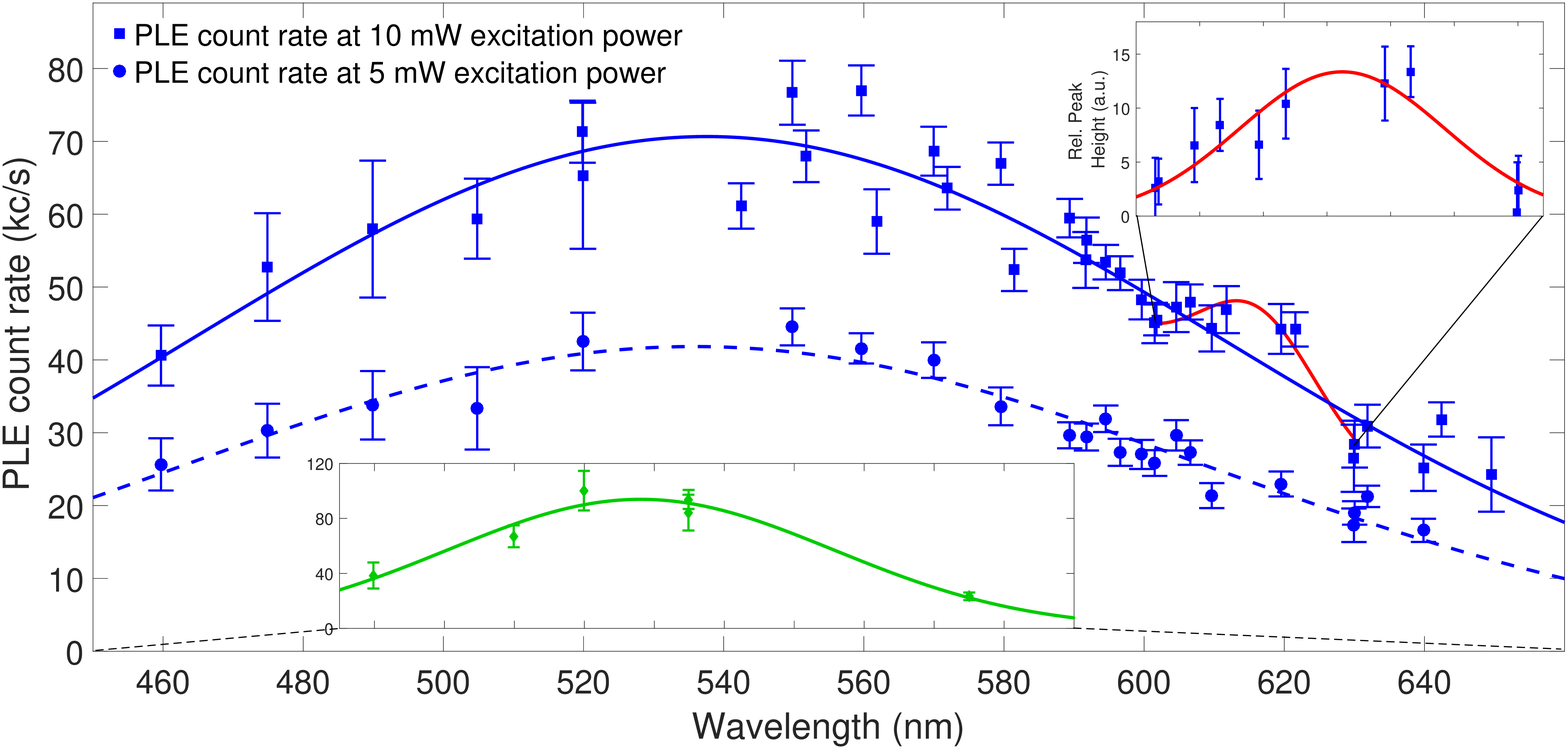}
	\end{minipage}
		\caption{\textbf{(a)} Photoluminescence spectrum of an ensemble of few SiV$^{-}$ centers taken at a pump power of $\approx 3 \, \text{mW}$ at $542 \, \text{nm}$ excitation wavelength. Zero-phonon line emission at $\approx \! 738 \, \text{nm}$ and phonon sideband up to $\approx 850 \, \text{nm}$ show typical characteristics of the SiV$^{-}$ color center at room temperature and no sign of other defects are visible. Using a Gaussian regression we obtain a FWHM of $5.3 \, \text{nm}$ (ZPL) and a DWF of $0.67$. \textbf{(b)} PLE spectra of two different SiV$^{-}$ ensembles for excitation wavelengths between $460 \, \text{nm}$ and $650 \, \text{nm}$. For the first ensemble we took PLE spectra at $5 \, \text{mW}$ (blue, dashed, circles) and $10 \, \text{mW}$ (blue, solid, squares) excitation power and for the second ensemble at $10 \, \text{mW}$ (green data in the inset). The dipole-allowed transition from $^{2}E_{g}$ to $^{2}A_{2u}$ is fit by Gaussian function with center wavelength at $536 \, \text{nm}$ and FWHM of $105 \, \text{nm}$ for the first ensemble and FWHM of $78 \, \text{nm}$ for the second ensemble. At $10 \, \text{mW}$ excitation power an additional peak at $616 \, \text{nm}$ with FWHM of $11.4 \, \text{nm}$ is observed, which is fitted with a Gaussian function after subtracting the background caused by the broad $^{2}A_{2u}$-resonance and setting the lowest datapoint to zero (see inset in upper right corner). For fitting of the broad $^{2}A_{2u}$-resonance these datapoints were excluded. \textbf{(c)} Excitation powerdependence of the photoluminescence signal of a SiV$^{-}$ ensemble at four different wavelengths. Data at $\lambda = 550 \, \text{nm}$ and $\lambda = 630 \, \text{nm}$ follow simplified saturation law (\ref{eqn:saturation}), while data at $\lambda = 590 \, \text{nm}$ and $\lambda = 610 \, \text{nm}$ clearly deviates from the saturation law at higher excitation power. The saturation curves for $\lambda = 550 \, \text{nm}$ and $\lambda = 630 \, \text{nm}$ are fitted using equation (\ref{eqn:saturation}) leading to saturation count rates of $204 \pm 24 \, \text{kcounts/s}$ and $57 \pm 5.4 \, \text{kcounts/s}$ and saturation powers of $19 \pm 3.6 \, \text{mW}$ and $10.4 \pm 1.5 \, \text{mW}$. For $\lambda = 590 \, \text{nm}$ and $\lambda = 610 \, \text{nm}$ we added saturation behavior derived from (\ref{eqn:saturation}) as guide to the eye (yellow, orange dotted lines) for increasing deviation from experimental data with increasing excitation power.}  
	\label{fig:SiV}
\end{figure}

An ensemble of few SiV$^{-}$ centers was studied in a $\{111\}$ diamond plate obtained by laser-slicing and polishing of a HPHT-grown type-IIa crystal. High-purity, $^{12}$C-enriched ($99.998\, \%$), homoepitaxial diamond film was deposited using a microwave plasma-assisted chemical-vapor-deposition (MPCVD) apparatus \cite{teraji2015homoepitaxial}. The total gas pressure, microwave power, methane concentration ratio ($\text{CH}_{4}$/total flow), oxygen concentration ratio ($\text{O}_{2}$/total flow), growth duration and substrate temperature employed were $120 \, \text{Torr}$, $1.4 \, \text{kW}$, $1 \, \%$, $0.2 \, \%$, $17 \, \text{h}$, and $990-1070 \, ^{\circ}\text{C}$, respectively. The 8N-grade $^{12}$C-enriched ($99.999 \, \%$) methane was used as a source gas. After the growth of non-doped layer, Si-doped layer was overgrown for $2 \, \text{h}$ by placing a SiC plate near the substrate resulting in a thickness of $3.8 \, \mu\text{m}$.\cite{rogers2014electronic, teraji2015homoepitaxial}. \\
SiV$^{-}$ and NV$^{-}$ center are formed at a few microns below the diamond surface. When using a $715 \, \text{nm}$ longpass filter in the detection path we reach count rates between $300 \, \text{kcounts/s}$ and $400 \, \text{kcounts/s}$ for several SiV$^{-}$ ensembles with $\approx \! 10 \, \text{mW}$ of $542 \, \text{nm}$ excitation power at the microscope objective, corresponding to around 10 SiV$^{-}$ centers in the detection volume. When using a $740/10 \, \text{nm}$ bandpass filter we obtain between $70 \, \text{kcounts/s}$ and $90 \, \text{kcounts/s}$ for the same spots. In the following we present measurement data for two different SiV$^{-}$ ensembles. \\
The fluorescent spots are first examined using spectrometer measurements. A photoluminescence spectrum of the first spot taken at an excitation power of $\approx \! 3 \, \text{mW}$ at wavelength of $542 \, \text{nm}$ laserlight is shown in figure \ref{fig:SiV}(a). We use a $600 \, \text{nm}$ longpass filter to cut out the laser emission. Typical emission characteristics of the SiV$^{-}$ color center at room temperature with a strong ZPL emission at $\approx \! 738 \, \text{nm}$ and a phonon sideband, which extends to $\approx \! 850 \, \text{nm}$ is observed. The Debye-Waller (DW) factor, defined as the fraction of fluorescence emitted into the ZPL compared to the total emission of the defect center is calculated, by using a Gaussian regression with $5.3 \, \text{nm}$ FWHM linewidth for the ZPL emission. As an outcome we find the DW factor to be $0.67$ which is comparable to earlier reports on the optical properties of the SiV$^{-}$ center \cite{dietrich2014isotopically}. No significant contribution from other optically-active defect centers within the detection volume of our confocal microscope and within the recorded spectral window are observed. 
Next, we perform PLE measurements by recording the emitted photons employing single photon counting modules and a $740/10 \, \text{nm}$ bandpass filter in the detection path while tuning the excitation wavelength over a wide range from $460 \, \text{nm}$ to $650 \, \text{nm}$ and keeping the excitation power constant. 
The resulting PLE spectra for two different excitation powers of $5 \, \text{mW}$ and $10 \, \text{mW}$ are shown in figure \ref{fig:SiV}(b), as well as a second data set for a different ensemble at $10 \, \text{mW}$ excitation power in the inset. Both ensembles show similar PLE characteristics with a very broad PLE signal distributed over the entire measured wavelength range. We fit a Gaussian lineshape to the measured data inferring the excitation maximum at around $536 \pm 5 \, \text{nm}$, with a FWHM linewidth of $105 \pm 9 \, \text{nm}$ for the first ensemble and $78 \pm 26 \, \text{nm}$ for the second ensemble. By comparison with theory we attribute the broad PLE resonance to a dipole-allowed transition from $^{2}E_{g}$ to $^{2}A_{2u}$ as elaborated in detail in section \ref{sec:el_struct}. Furthermore, we take PL spectra at most of the measured excitation wavelengths and do not observe a significant change in the fluorescence spectrum. Independent on excitation wavelength there is always a rapid decay process present leading to a fast decay, fast compared to the optical lifetime of the $^{2}A_{2u}$-state, into the $^{2}E_{u}$-state and a subsequent optical transition into the ground state $^{2}E_{g}$ with the characteristic ZPL emission at $\approx \! 738 \, \text{nm}$. 
So far, excitation power is kept constant to obtain the wavelength-dependent PLE spectra. Next we study the excitation wavelength-dependent saturation behavior of the SiV$^{-}$ ensembles. 
Figure \ref{fig:SiV}(c) shows the measurement data at four different excitation wavelengths in the range between $550 \, \text{nm}$ and $630 \, \text{nm}$. 
Close to the maximum of the measured wavelength-dependent PLE curve at $536 \, \text{nm}$ we observe saturation behavior that we can fit with simplified model of the saturation law 
\begin{equation}
I = I_{\infty}(P/P_{s}) \cdot (1 + P/P_{s})^{-1},
\label{eqn:saturation}
\end{equation}
where $I_{\infty}$ is the saturation count rate and $P_{s}$ the saturation power. Exciting at $550 \, \text{nm}$ (green data in figure \ref{fig:SiV}(c)) we obtain a saturation count rate of $I_{\infty} = 204 \pm 24 \, \text{kcounts/s}$ and a saturation power of $I_{s} = 19 \pm 3.6 \, \text{mW}$ (green curve). 
We conclude that we are exciting an optical allowed transition with a transition resonance at $\approx \! 536 \, \text{nm}$. At constant excitation power the PLE count rate decreases with increasing detuning from resonance. For higher excitation wavelengths between $630 \, \text{nm}$ (red data in figure \ref{fig:SiV}(c)) and $650 \, \text{nm}$ we observe saturation behavior in agreement with the saturation law (\ref{eqn:saturation}) but with significantly reduced saturation count rate of $57 \pm 5.4 \, \text{kcounts/s}$ and saturation power of $10.4 \pm 1.5 \, \text{mW}$ (red curve). However, for excitation wavelengths of $590 \, \text{nm}$ and $610 \, \text{nm}$ (yellow and orange data in figure \ref{fig:SiV}(c)) we observe a significant change in the saturation behavior. For illustration we plot results from saturation law (\ref{eqn:saturation}) as guide to the eye. Significant deviation from the saturation law is observed with increasing discrepancy for increasing excitation power. No sign of saturation is visible for excitation powers up to $10 \, \text{mW}$. Furthermore, we observe a signature of a spectral feature at excitation power of $10 \, \text{mW}$, marked as red solid line in the PLE spectrum of figure \ref{fig:SiV}(b). We fit the additional peak in the PLE spectrum with a Gaussian regression after subtracting the background caused by the broad $E_{g} \rightarrow A_{2u}$ resonance resulting in a peak position at $616 \pm 4 \, \text{nm}$ and a peak width of $11.4 \pm 1.7 \, \text{nm}$.
We interpret our experimental findings as the onset of driven parity-conserving transition at a peak position of $616 \pm 4 \, \text{nm}$. So far, we have mapped optical transitions which are parity-altering via single photon excitation. However, with increasing excitation power we hypothesize to start to probe also parity-conserving transitions, probably via two-photon excitation as it has been reported recently in reference \cite{higbie2017multiphoton}. Our excitation laser system is simultaneously tuned between $900 \, \text{nm}$ and $1300 \, \text{nm}$ by tuning the OPO and, at the same time, frequency-doubled light is generated in the visible range between $450 \, \text{nm}$ and $650 \, \text{nm}$ via second harmonic generation. For our measurement at $10 \, \text{mW}$ excitation power and with excitation wavelength of $620 \, \text{nm}$ we extrapolated that $\approx 10 \, \mu\text{W}$ of $1240 \, \text{nm}$ light is present in front of the objective which could be sufficient to probe the onset of parity-conserving transitions via two-photon excitation. 
According to our theory, the lowest energy parity-conserving excited state can be described by promoting an electron from the valence band edge to the $e_g$ defect level. The valence band edge will split in the presence of the defect to $a_{1g}$ and $e_g$ bands in the $\Gamma$-point of the Brillouin-zone, so they correspond to gerade bands. The corresponding excitation energy is smaller than the excitation energies related to the $a_{2u}$ bands because the $a_{2u}$ bands lie deeper in the valence band. 
Our observed signature at $616 \, \text{nm}$ is in good agreement with recent results obtained from two-photon excited fluorescence spectra reported in reference \cite{higbie2017multiphoton} where the parity-conserving optical transitions are probed via two-photon excitation. The threshold of two-photon absorption can map the ionization energy minus the binding energy of the exciton to the SiV$^{-}$ defect (gerade-to-gerade transition) whereas single-photon excitation can predominantly excite higher energy states.\\
Furthermore, we want to note that the observed spectral feature happens very close to the expected SiV$^{2-}$ charge transition which is estimated to be resonant at around $2.1 \, \text{eV}$ ($\approx \! 590.4 \, \text{nm}$) (see \cite{gali2013ab} and section~\ref{sec:el_struct}). For in-depth interpretation of the ionization process further investigations at higher excitation powers and with single-shot charge state detection are necessary \cite{aslam2013photo}.   

\cleardoublepage 
\newpage
\newpage

\section{Investigation of ensemble of few GeV$^{-}$ center}

\begin{figure}[htbp]
	\begin{minipage}{0.49\textwidth}
			\textbf{(a)} \\
			\includegraphics[width=\textwidth]{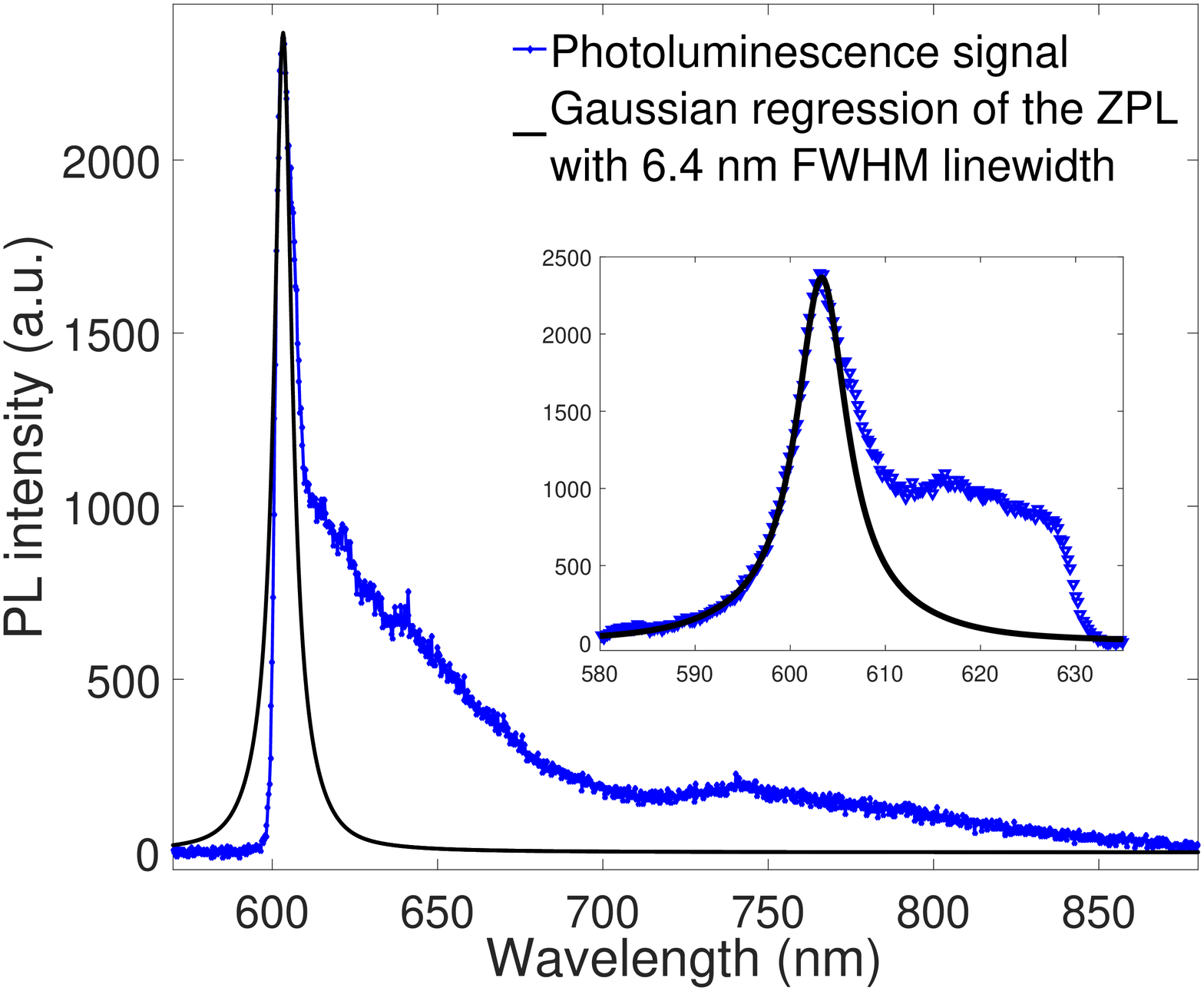}
	\end{minipage}
	\begin{minipage}{0.49\textwidth}
			\textbf{(c)} \\
			\includegraphics[width=\textwidth]{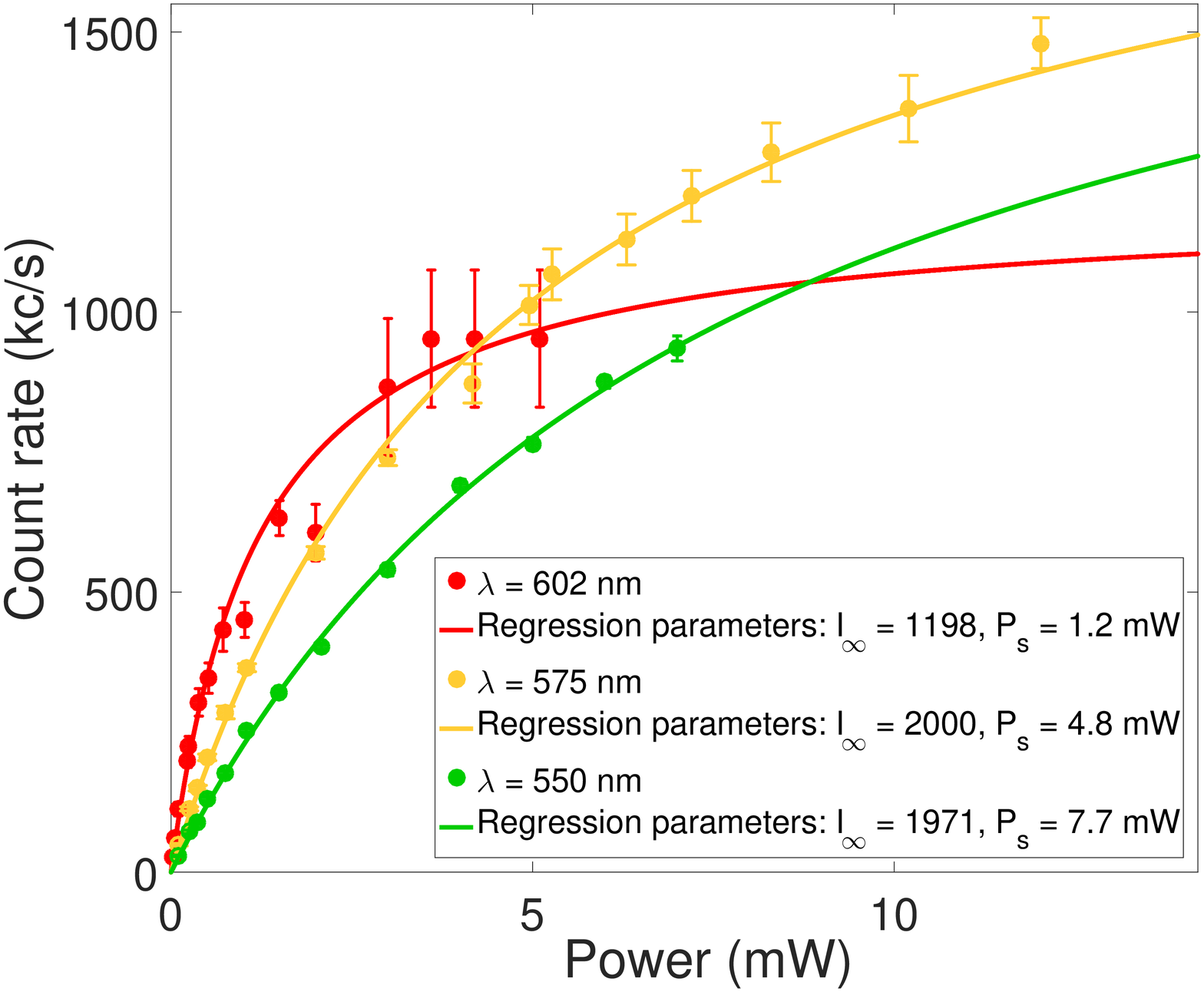}
	\end{minipage}
	\begin{minipage}{\textwidth}
			\textbf{(b)} \\
			\includegraphics[width=\textwidth]{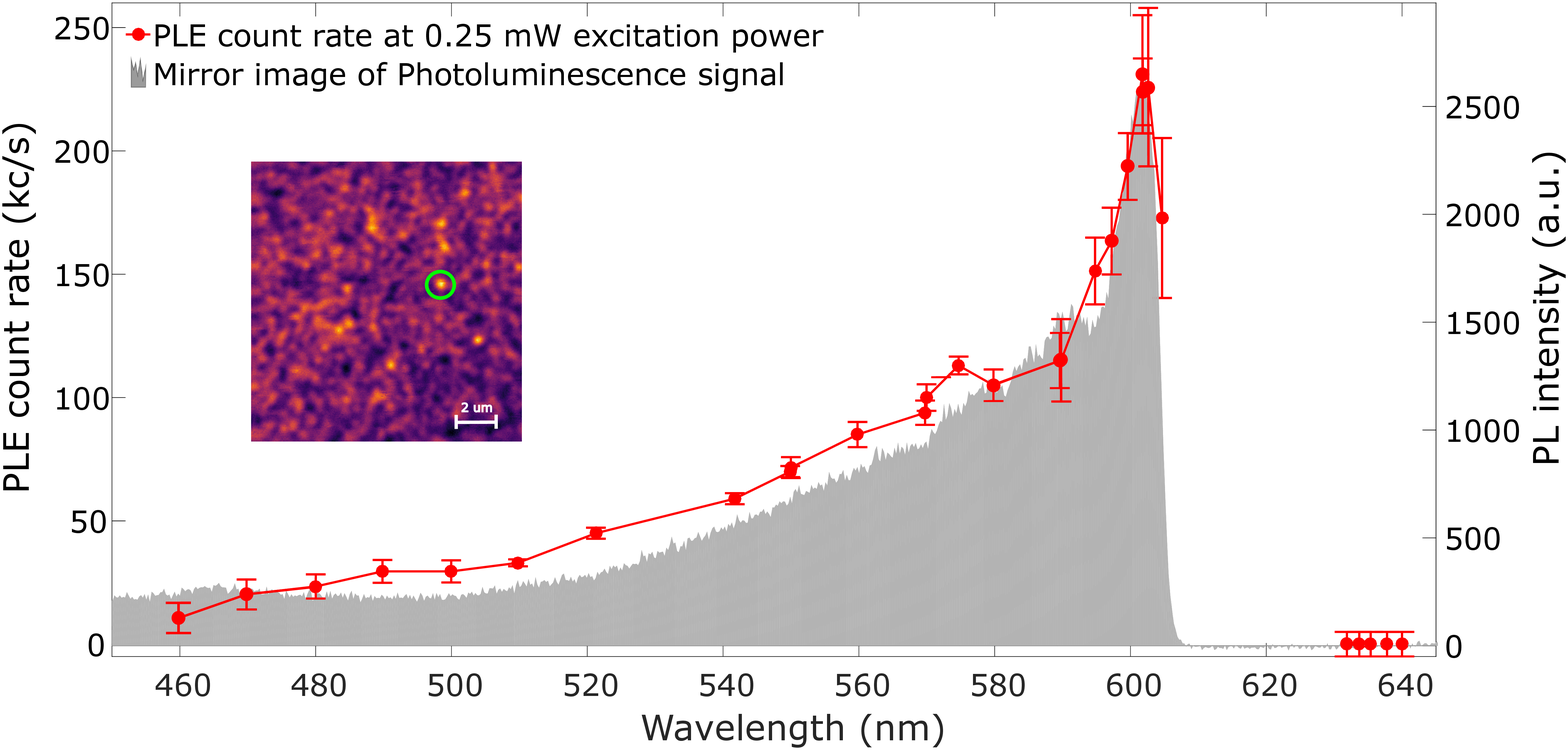}
	\end{minipage}
		\caption{\textbf{(a)} Photoluminescence spectrum of the examined GeV$^{-}$ ensemble marked in the inset of figure \ref{fig:GeV}(b) at an excitation power of $\approx \! 0.5 \, \text{mW}$ at $542 \, \text{nm}$ excitation wavelength. We identified fluorescence of GeV$^{-}$ center with zero-phonon line at $\approx 602 \, \text{nm}$ and a $200 \, \text{nm}$ broad phonon sideband. In the main image the filter cutoff at $600 \, \text{nm}$ is visible, which supresses part of the ZPL emission. The inset shows the whole ZPL when using a $602/50 \, \text{nm}$ bandpass filter in the detection path, enabling a Gaussian regression of the ZPL with a FWHM linewidth of $\approx \! 6.4 \, \text{nm}$ and the calculation of the DW factor of $\geq \! 0.26$. \textbf{(b)} PLE spectrum of the GeV$^{-}$ ensemble for excitation wavelengths between $460 \, \text{nm}$ and $640 \, \text{nm}$ at $0.25 \, \text{mW}$ excitation power. We find an excitation maximum at the ZPL wavelength of $\approx \! 602 \, \text{nm}$ with count rates of almost $250 \, \text{kcounts/s}$. In addition we see a decay of the excitation probability when tuning away from resonance on the blue wavelength side. Between $460 \, \text{nm}$ and $605 \, \text{nm}$ the PLE spectrum represents the mirror image of the photoluminescence spectrum, which is sketched by the grey shaded area. Note the scaling of the two y-axes and the inverted x-axis of the PL spectrum to achieve matching. \textbf{Inset.} Confocal image of a $10 \, \mu\text{m}\,\times\,10 \, \mu\text{m}$ area of the high density GeV sample with the investigated GeV$^{-}$ ensemble encircled. \textbf{(c)} Excitation powerdependence of the photoluminescence signal of GeV$^{-}$ ensemble at three different wavelengths. We see saturation behavior for quasi resonant excitation of the ZPL at $602 \, \text{nm}$ as well as for excitation via sideband ($\lambda = 550 \, \text{nm}$ and $\lambda = 575 \, \text{nm}$). The saturation curves are fitted using equation (\ref{eqn:saturation}) leading to a saturation count rate of $\approx 1200 \, \text{kcounts/s}$ at $602 \, \text{nm}$ and almost twice as high for off-resonant excitation. The saturation power increases about a factor of four (six) comparing excitation with $\lambda = 602 \, \text{nm}$ and $\lambda = 575 \, \text{nm}$ ($\lambda = 550 \, \text{nm}$).}  
	\label{fig:GeV}
\end{figure}

In order to study optical properties of the negatively charged germanium-vacancy center we investigate a high-density GeV sample fabricated by ion implantation with a implantation dose of $3.5 \cdot 10^{10} \, \text{cm}^{-2}$ and subsequent anneal at $800 \, ^{\circ}$C in a IIa-type diamond substrate. Around $60 \, \text{nm}$ below the surface a bright layer is formed showing typical GeV$^{-}$ emission characteristics. Several bright spots can be identified as GeV$^{-}$ ensembles with up to ten particles within the detection volume. We measure a maximum of $300 \, \text{kcounts/s}$ for GeV$^{-}$ ensemble marked in figure \ref{fig:GeV}(d) with an excitation power of $\approx \! 0.5 \, \text{mW}$ at the microscope objective and $542 \, \text{nm}$ excitation wavelength. In the detection path we use a $600 \, \text{nm}$ longpass filter. Due to the high density of GeV$^{-}$ centers in the sample we obtain a background fluorescence of $100 \, \text{kcounts/s}$ with a sample background also showing GeV$^{-}$ spectra. \\
Figure \ref{fig:GeV}(a) shows the PL spectrum of the observed GeV$^{-}$ ensemble recorded with the spectrometer. The ZPL at $\approx \! 602 \, \text{nm}$ is visible but partially suppressed by the filter cutoff at $600 \, \text{nm}$. By using a $605/50 \, \text{nm}$ bandpass filter in the detection path we can map the whole ZPL and fit it with a Gaussian lineshape with a FWHM of $6.4 \, \text{nm}$ to calculate DW factor of $\geq \! 0.26$ (see inset in figure \ref{fig:GeV}(a)). 
PLE measurements of the GeV$^{-}$ ensemble with an excitation power of $\approx \! 0.25 \, \text{mW}$ in the wavelength range from $460 \, \text{nm}$ to $640 \, \text{nm}$ are shown in figure \ref{fig:GeV}(b). We use $600 \, \text{nm}$ longpass filter for excitation wavelengths below $600 \, \text{nm}$ and $610 \, \text{nm}$ longpass filter for excitation wavelengths between $600 \, \text{nm}$ and $605 \, \text{nm}$ leading there to larger error bars due to reduced count rates. The overall PLE count rates are renormalized taking fractional filter transmission into account. 
The maximum excitation with a count rate of $\approx 250 \, \text{kcounts/s}$ can be obtained for resonantly driving the ZPL at $\approx \! 602 \, \text{nm}$. In the wavelength range between $460 \, \text{nm}$ and $605 \, \text{nm}$ the PLE spectrum is in good agreement with the mirror image of the photoluminescence spectrum flipped at precisely the central ZPL emission frequency (shown in figure \ref{fig:GeV}(b) as the grey shaded plot). On the blue wavelength side we see a decay of the excitation probability, which drops almost to zero at $460 \, \text{nm}$. For excitation wavelengths between $630 \, \text{nm}$ and $640 \, \text{nm}$ no excitation of the GeV$^{-}$ ensemble is possible. In analogy to the measurements on the SiV$^{-}$ ensembles we do not observe significant change of the PL spectrum over the whole excitation wavelength range. \\
To investigate the excitation power-dependence of the fluorescence of the GeV$^{-}$ ensemble we present measurement data of the fluorescence count rate as a function of excitation power for three different excitation wavelengths in figure \ref{fig:GeV}(c). The saturation behavior over the whole excitation wavelength range is in good agreement with above mentioned model of the saturation law. Using equation (\ref{eqn:saturation}) we calculate saturation curves for the three data sets in figure \ref{fig:GeV}(c). As a result we get a saturation count rate of $\approx 1200 \, \text{kcounts/s}$ at $602 \, \text{nm}$ and almost twice as high saturation count rates for excitation via sideband ($\lambda = 575 \, \text{nm}$ and $\lambda = 550 \, \text{nm}$). We justify the increased saturation count rate with a reduced population in the groundstate for strong off-resonant excitation while the population in the groundstate remains $50 \, \%$  for strong resonant excitation. The reduced excitation probability for sideband excitation is demonstrated by the increase of the saturation power at $\lambda = 575 \, \text{nm}$ ($\lambda = 550 \, \text{nm}$) excitation by a factor of four (six).
It is important to  note that the $\text{GeV}^{-} \, \rightarrow \, \text{GeV}^{2-}$ charge transition level is at around $2.5 \, \text{eV}$ according to our calculations. This means that PLE measurements on the second band should compete with the ionization process, but in the case of GeV$^{-}$ this competition is less severe than that for SiV$^{-}$.

\cleardoublepage 
\newpage
\newpage

\section{Electronic level structure of the SiV$^{-}$ and GeV$^{-}$ color center}
\label{sec:el_struct}

\begin{figure}[htbp]
	\begin{minipage}{\textwidth}
			\begin{flushleft}
			\textbf{(a)}
			\end{flushleft}
			\centering
			\includegraphics[width=0.59\textwidth]{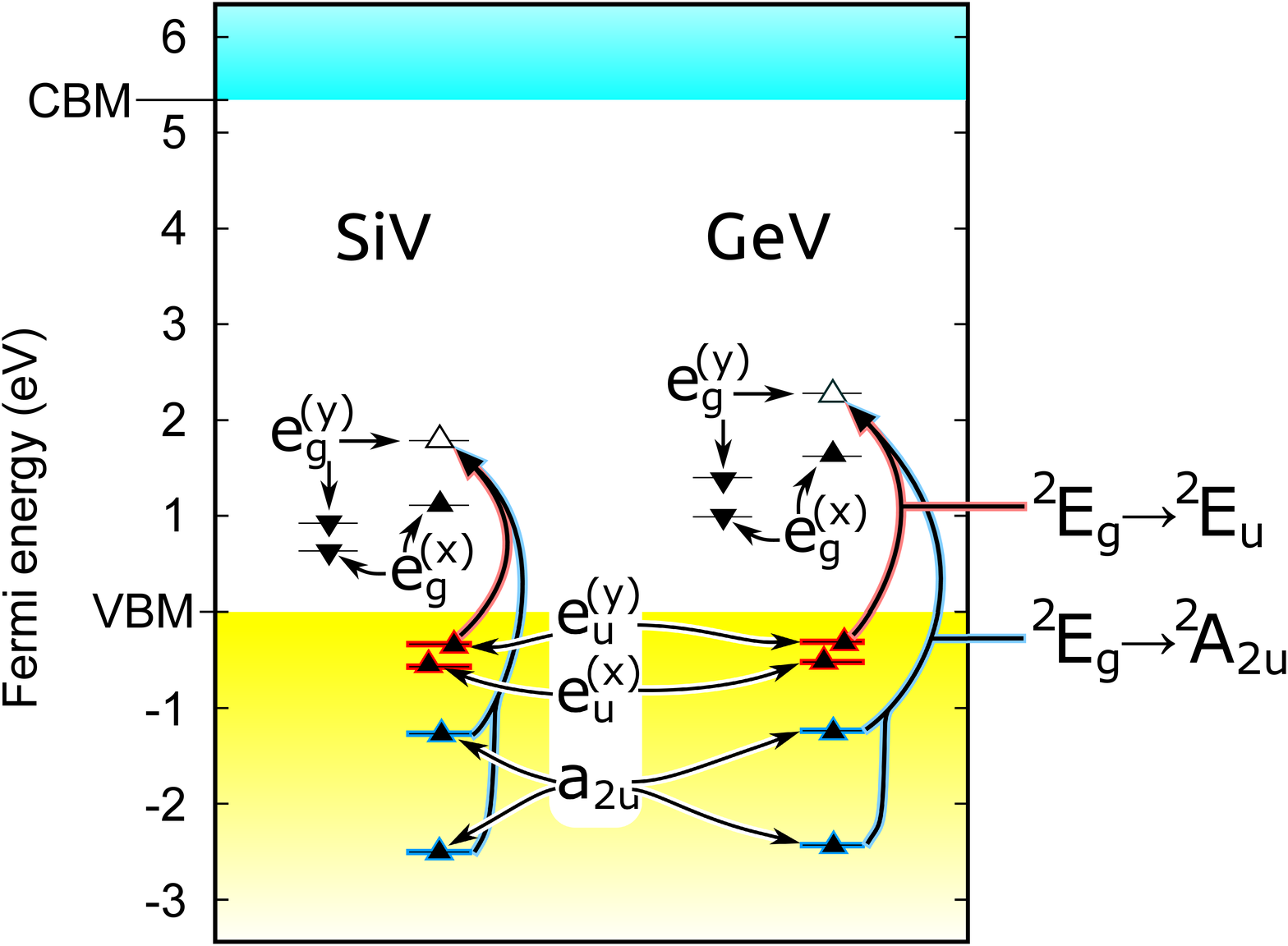}
	\end{minipage}
	\begin{minipage}{\textwidth}
			\begin{flushleft}
			\textbf{(b)}
			\end{flushleft}
			\centering
			\includegraphics[width=0.8\textwidth]{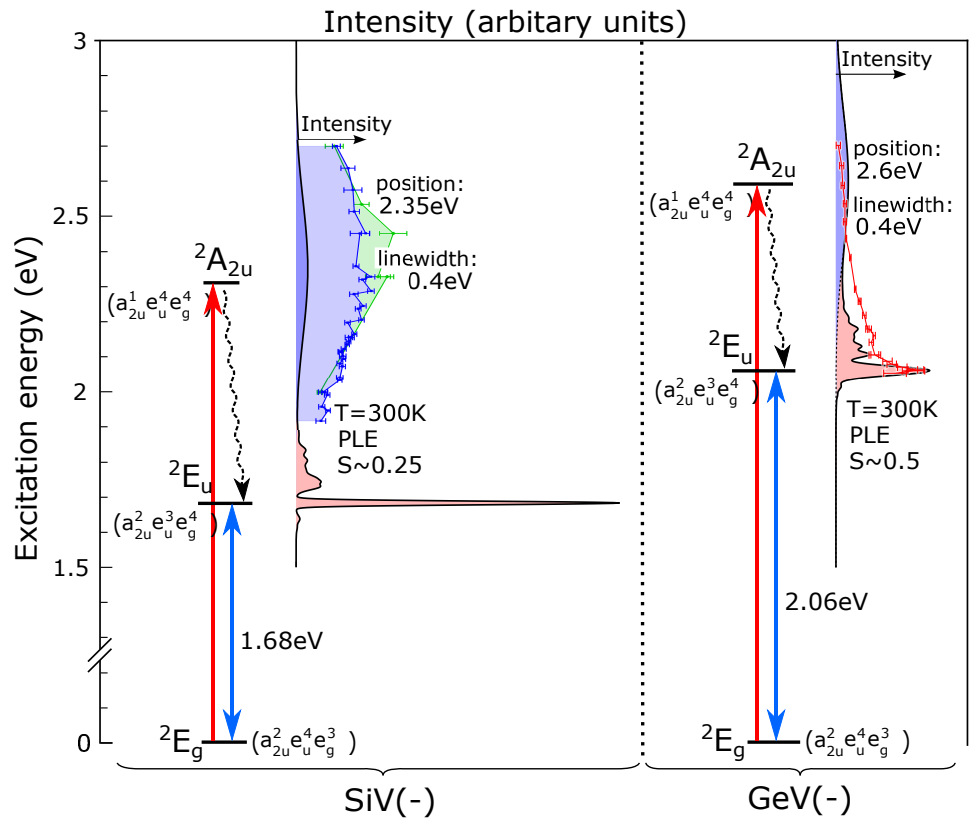}
	\end{minipage}
		\caption{\textbf{(a)} Electronic configuration for the ground state ($^{2}E_{g}$) as calculated by HSE06 functional in a 512-atom supercell. The energy of the valence band maximum is aligned to $0 \, \text{eV}$. The optical transition associated with the first excited state ($^{2}E_{u}$) and the second excited state ($^{2}A_{2u}$) are depicted by red and blue inclined arrows, respectively. The $e_{u}$ Kohn-Sham state is getting localized when a hole is induced on it according to constraint DFT calculations (see also ref~\cite{gali2013ab}). This does not hold for $a_{2u}$ bands. \textbf{(b)} Developed model for the electronic level structure of the SiV$^{-}$ and GeV$^{-}$ center combining the experimental data from PLE spectroscopy and the theoretical calculations. $S$ is the calculated Huang-Rhys factor of the first band. The 'position' and $\sigma$ refers to the derived parameters of the Gaussian function that represents the second band in PLE. For SiV$^{-}$ defect the experimental PLE results are plotted for the second band (c.f. figure \ref{fig:SiV}(b)). See text and the Supplementary Materials for explanation.}  
	\label{fig:Levelscheme}
\end{figure}

Finally, we develop a consistent physical model of the electronic level structure of the SiV$^{-}$ and GeV$^{-}$ center. Figure \ref{fig:Levelscheme}(a) depicts the electronic structure for the ground state of SiV$^{-}$ and GeV$^{-}$. One can see that the electronic structure of the two defects are very similar except that the in-gap $e_g$ level lies about $0.3 \, \text{eV}$ higher for GeV$^{-}$ than that for SiV$^{-}$. In our calculations we allow a small static distortion to $C_{2h}$ symmetry. This results in a small splitting of the degenerate $e_g$ and $e_u$ states. We disregard spin-orbit coupling in simulations due to different energy scale with respect to optical transition energy. The $e_u$ states are closely resonant with the valence band maximum (VBM). We already showed in our previous publication for SiV$^{-}$ by constraint DFT calculation \cite{gali2013ab} that if a hole is induced for this $e_u$ state then it becomes localized. Thus, the transition to the first excited state ($E_{u}$) results in a sharp zero-phonon line (ZPL) transition. The participation of the phonons in this process can be calculated for this transition by using different level of approximations in the electron-phonon coupling \cite{Londero_2016}. We find that the same effect holds for the first excited state of GeV$^{-}$ just its ZPL energy lies about $0.3 \, \text{eV}$ higher.

We find that a broad $a_{2u}$ band is developed below the $e_u$ level. The $a_{2u}$ defect level is strongly mixed with the diamond bands that creates this broad $a_{2u}$ band. This $a_{2u}$ band emerges about $1.3 \, \text{eV}$ below VBM, and is about $1.5 \, \text{eV}$ broad. If we promote an electron from any of these $a_{2u}$ states to the in-gap $e_g$ state then the resulted hole will never be localized. This means that the absorption spectrum arising from these $a_{2u}$ band will be broad even when just pure electronic transitions are considered. We find the optically forbidden $a_{1g}$ band at least $1 \, \text{eV}$ deeper in energy in the valence band than the $a_{2u}$ band. Since the $E_g \rightarrow A_{2u}$ optical transition is optically allowed we calculated the absorption spectrum at the ground state geometry at DFT PBE level in a large 4096-atom supercell but using the HSE06 Kohn-Sham energies in the estimate of the optical transition energies.  In this context we measure the excitation energies coming from the $a_{2u}$ band from the $e_u$ level as a reference that we associate with the $E_u \leftrightarrow E_g$ ZPL energy. This means that we implicitly assume a similar Stokes-shift in the $E_g \rightarrow A_{2u}$ transition as that in the $E_{u} \leftrightarrow E_g$ transition. We found that the integrated absorption of the $E_g \rightarrow A_{2u}$ transitions is almost equal to that of the $E_{u} \leftrightarrow E_g$ transition. 

Having all of these data in our hand, a simulated absorption spectrum is generated shown in Figure \ref{fig:Levelscheme}(b). The calculated HSE06 ZPL energies of the $E_u \leftrightarrow E_g$ transition are $1.72 \, \text{eV}$ and $2.15 \, \text{eV}$ for the SiV$^{-}$ and GeV$^{-}$ defects, respectively, to be compared to the experimental values at $1.68 \, \text{eV}$ and $2.06 \, \text{eV}$, respectively. The $E_u \leftrightarrow E_g$ ZPL energy is aligned to the experimental value for the considered defects in Figure \ref{fig:Levelscheme}(b), for the sake of clarity and direct comparison with the experimental data. The second band in the absorption spectrum that we associate with the optical transition from the $a_{2u}$ band is generated as follows. We assume that the 4096-atom supercell calculation is still not fully convergent to describe the mixture of $a_{2u}$ band with the diamond bands, so the calculated spectrum is still too discrete (see Supplementary Materials). Therefore, we calculate the averaged value of the $a_{2u}$ band as $\left\langle x\right\rangle =\left(\int xf\left(x\right)dx\right)/\left(\int f\left(x\right)dx\right)$ and its standard deviation as $\sigma^{2}=\left(\int\left(x-\left\langle x\right\rangle \right)^{2}f\left(x\right)dx\right)/\left(\int f\left(x\right)dx\right)$ from the 4096-atom supercell. The $f(x)$ is the trace of the frequency dependent dielectric matrix with optical transitions only for $a_{2u}$ band. Similarly we calculate the averaged position of the $e_u$ band, then add the energy difference between the $e_u$ and $a_{2u}$ into the experimental ZPL energy of the $E_g \leftrightarrow E_u$ transition. Then for the sake of simplicity we plot a Gaussian function with $\sigma$ standard deviation at the given position.  The total integrated absorption of the Gaussian function follows the total integrated absorption associated with $E_g \rightarrow A_{2u}$ transitions as obtained in the 4096-atom calculation. The results of this fitting procedure provide the central position (highest intensity) of the second band in the absorption, and the broadening. We find that the central position of the second band in the absorption spectrum is at $2.29 \, \text{eV}$ and $2.72 \, \text{eV}$ for SiV$^{-}$ and GeV$^{-}$ defects, respectively. The calculated value for SiV$^{-}$ agree nicely with the experimental data. However, the calculated broadening of the second band is larger than that in the experiment. We argue that the broadening of the second band in the absorption is overestimated because the electron-hole interaction was not taken into account in our method that should somewhat localize the $a_{2u}$ band. Nevertheless, our calculations prove that allowed optical transitions should occur in the energy region where the second band in the PLE spectrum was observed. For GeV$^{-}$ defect the calculated absorption spectrum is a bit more complex. The phonon sideband of the usual $E_{u} \leftrightarrow E_{g}$ transition overlaps almost completely with the very broad and low intensity $a_{2u}$-band to $E_{g}$ optical transitions, therefore, it is much more difficult to observe the presence of this second band. Nevertheless, the calculated broadening of the second band is still somewhat overestimated due to the neglect of electron-hole interaction in the calculated spectrum that results in some probability of optical transition at shorter wavelengths than that was observed experimentally. We further note that the polarization of the second band in the PLE spectrum \cite{rogers2014electronic} agrees well with our model as $E_g \rightarrow A_{2u}$ optical transitions are only allowed by photons polarized perpendicular to the symmetry axis of the defect.

\cleardoublepage 
\newpage
\newpage

\section{Conclusion}

Our experimental investigation on the excitation of an ensemble of SiV$^{-}$ centers and the direct comparison with theoretical modeling of the imaginary dielectric function suggest a dipole-allowed transition resonant at $536 \, \text{nm}$ from $^{2}E_{g}$-state to $^{2}A_{2u}$-state where the corresponding $a_{2u}$-level lies deeply inside the diamond valence band leading to broadening of the transition with FWHM of $105 \, \text{nm}$. We observe a change of saturation behavior in the wavelength range between $590 \, \text{nm}$ and $610 \, \text{nm}$. Together with a spectral peak at $616 \, \text{nm}$ with a FWHM of $11.4 \, \text{nm}$ arising at higher excitation power of $10 \, \text{mW}$ we interpret our data as probing of a parity-conserving transition from the valence band edge, split in the presence of the defect to $a_{1g}$ and $e_g$ bands in the $\Gamma$-point of the Brillouin-zone, corresponding to gerade bands. We further note that the change in saturation behavior happens close to the SiV$^{2-}$ charge transition at around $2.1 \, \text{eV}$, which was proposed in \cite{gali2013ab}. \\
For an ensemble of GeV$^{-}$ centers we find the PLE spectrum in good agreement with the mirror image of the photoluminescence spectrum. Theoretically we propose a similar electronic level structure for GeV$^{-}$ as compared to SiV$^{-}$, resulting in a second excited state $^{2}A_{2u}$ at $2.6 \, \text{eV}$. Experimentally we do not confirm the presence of this state due to the low intensity and large width of the $a_{2u}$-band, which overlaps with the phonon sideband of the $E_{u} \leftrightarrow E_{g}$ transition. PLE measurements at low temperature could help to get better resolved PLE spectra for GeV$^{-}$ as has recently been reported in \cite{ekimov2017anharmonicity} between $2.05$ and $2.25 \, \text{eV}$. Extending the measurements to shorter wavelength could shed light on the existence of a weak maximum around the calculated $2.6 \, \text{eV}$ due to $E_{g} \rightarrow A_{2u}$ transition. \\
\\ 
An alternative interpretation of the electronic level structure of the two centers is based on a forbidden transition with phonon-allowed band to the $a_{1g}$-state as proposed in \cite{rogers2014electronic}. Our DFT results rather imply that broadening of the second band in the PLE spectrum is due to the optically active $a_{2u}$ bands, and no sharp optical signals are expected from band to in-gap level type of photo-excitation. The $a_{2u}$ level is supposed to lie above the $a_{1g}$ level and thus the spectroscopic feature closest in energy to the ZPL would be more likely due to $a_{2u}$

\cleardoublepage 
\newpage
\newpage

\section{Supplementary Material}

We provide additional information about the calculated optical spectrum that is plotted in figure \ref{fig:Levelscheme}(b) of the main text. The calculations involve large supercell modeling and consideration about the limitation of the applied Kohn-Sham DFT functional. The corresponding raw data and optical spectra are shown in figure \ref{fig:supp}, and the ab-initio derivation of the corresponding parameters applied in producing figure \ref{fig:Levelscheme}(b) are described in the corresponding figure caption.
\begin{center}
	\begin{minipage}{\textwidth}
			\begin{flushleft}
			\textbf{(a)}
			\end{flushleft}
			\centering
			\includegraphics[width=\textwidth]{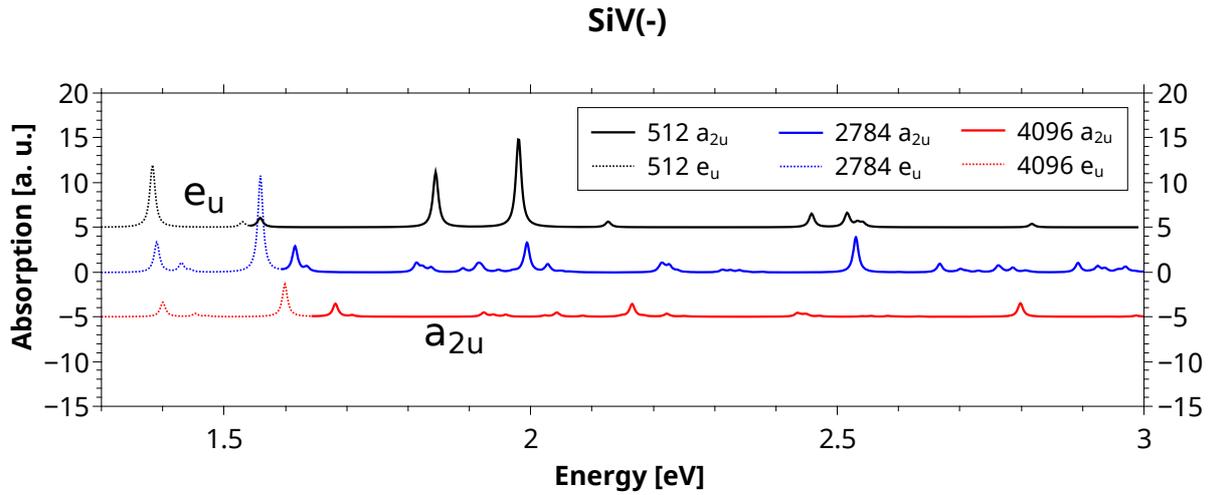}
	\end{minipage}
	\begin{minipage}{\textwidth}
			\begin{flushleft}
			\textbf{(b)}
			\end{flushleft}
			\centering
			\includegraphics[width=0.85\textwidth]{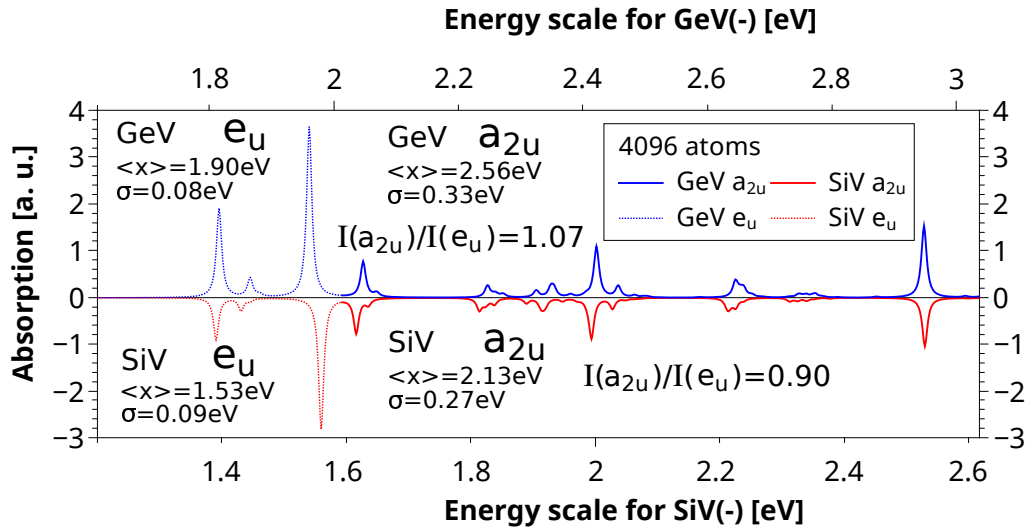}
	\end{minipage}
		\captionof{figure}{\textbf{(a)} Imaginary part of the dielectric function for the SiV$^{-}$ system in supercells with various sizes that is associated with the optical transitions. The spectrum was calculated within PBE DFT functional without taking into account the electron-hole (exciton) interaction and ionic relaxation effects, thus the calculated optical transition energies should not be compared to the experimental data. The values in the inset label the number of atoms in the perfect supercell. The structure of the optical transitions is the same for different supercell sizes but the calculated spectra are smeared more by increasing the size of the diamond supercell. The $e_u$ and $a_{2u}$ labels the holes that are created upon optical excitation. When the hole is created in the $a_{2u}$ bands then the spectrum is already spread in 512-atom supercell, and then spreads out even more in enlarged supercells. The $e_u$ hole splits in 2784-atom and 4096-atom supercells that we attribute to the PBE DFT approximation and the neglect of excitonic effects. In the constraint HSE06 DFT approach the $e_u$ hole is always localized. Nevertheless, the total absorption cross section associated with the $e_g \rightarrow e_u$ transitions within PBE DFT approach should be similar to that obtained by HSE06 DFT. \\
\textbf{(b)} Imaginary part of the dielectric function for the SiV$^{-}$ and GeV$^{-}$ systems. The spectra were calculated in 4096-atom supercell within PBE DFT approach without excitonic effects. By aligning the energy scale of the calculated optical transition energies of the two defects one can see the similarity of the two systems. We define the peak position by calculating the first moment $\left\langle x\right\rangle$ and the standard deviation $\sigma$ of the optical spectrum for each hole, separately. The value of the first moment provides the peak position. First, we calculated this for the $e_u$ hole bands. The resultant $\sigma = 0.09 \, \text{eV}$ for this optical transition of the SiV$^{-}$ defect is overestimated because of the afore-mentioned approximations in the calculation. After getting the calculated energy position of the $e_u$ transitions we shifted the energy scale to match the calculated $e_u \rightarrow e_g$ peak with the experimental zero-phonon-line for each defect, that are, $1.68 \, \text{eV}$ for SiV$^{-}$, and $2.06 \, \text{eV}$ for GeV$^{-}$, respectively. These results in $+0.15 \, \text{eV}$ and $+0.16 \, \text{eV}$ rigid shifts for these defects, respectively. By applying these energy shifts also to the calculated peak positions of the $a_{2u}$ spectra one yields $2.29 \, \text{eV}$ and $2.72 \, \text{eV}$ for SiV$^{-}$ and GeV$^{-}$ defects, respectively. These data are depicted in Fig. \ref{fig:Levelscheme} of the main text. The calculated $\sigma$ of these $a_{2u}$ spectra are used to estimate the width of the optical spectrum associated with the $a_{2u}$ hole which is an overestimation due to the neglect of excitonic effects in the calculations.}
	\label{fig:supp}
\end{center}

\cleardoublepage 
\newpage
\newpage

\section*{Acknowledgements}

We thank Dr. Hitoshi Sumiya (Sumitomo Electric Industries, Ltd.) for providing the HPHT crystal for the SiV sample. This work was supported by PRESTO, JST (Grant Number JPMJPR16P2) and CREST, JST and in part by Japan Society for the Promotion of Science KAKENHI (No.26246001 and No. 15H03980). We further thank Lachlan Rogers for fruitful discussions. Experiments performed for this work were operated using the Qudi software suite. SH and AK acknowledge support of IQST. AK and AD acknowledge support of the Carl-Zeiss Foundation. AK acknowledges support of the Wissenschaftler-Rückkehrprogramm GSO/CZS and DFG. FJ acknowledges support of the DFG, BMBF, VW Stiftung and EU (ERC, DIADEMS). AG acknowledge EU DIADEMS funding.

\nocite{*}
\section*{References}
\bibliographystyle{unsrt}
\bibliography{PLE_spectroscopy_SiV_GeV}

\end{document}